\documentclass[aps,prl,twocolumn,nofootinbib,longbibliography]{revtex4-1}

\usepackage{amsfonts}
\usepackage{amsmath}
\usepackage{amssymb}
\usepackage{amsthm}
\usepackage{setspace}
\usepackage{fullpage}
\usepackage{graphicx}
\usepackage{url}
\usepackage{braket}
\usepackage{mathtools}
\usepackage[all]{xy}

\usepackage{color}
\usepackage{bm}

\usepackage{verbatim}
\usepackage{graphicx}

\usepackage{mathrsfs}

\newtheorem{ex}{Example}

\newcommand{\beq}{\begin{eqnarray}}
\newcommand{\eeq}{\end{eqnarray}}

\begin{document}

\title{Gravitational Solitons and Completeness}

\author{Jacob McNamara}
\affiliation{Jefferson Physical Laboratory, Harvard University, Cambridge, MA 02138, USA}

\begin{abstract}

We show that gravitational solitons naturally carry gauge charges beyond those of any local quantum field. The effect of these charged excitations is to break a non-invertible symmetry to its maximal group-like sub-symmetry. Taking these charges into account, we show that the Completeness Hypothesis follows from the breaking of the remaining group-like symmetry. We generalize this picture to an arbitrary semisimple tensor category of particle charges, showing that the charges of gravitational solitons form the adjoint subcategory. We discuss a further generalization involving the charges of extended objects.

\end{abstract}

\maketitle

An essential feature of quantum gravity is the fluctuation of the topology of spacetime. While these fluctuations are suppressed at macroscopic scales, they become large at the Planck scale \cite{Hawking:1978pog}. As a result, the UV structure of quantum gravity is qualitatively different from that of a continuum quantum field theory, in which space and time are featureless at arbitrarily short distances. Understanding the physical consequences of this difference, especially on IR physics, is the goal of the Swampland Program.\footnote{For a review of the Swampland Program, see \cite{Brennan:2017rbf, Palti:2019pca, vanBeest:2021lhn, Grana:2021zvf}.}

Two basic swampland conditions, expected to be true in all consistent theories of quantum gravity, are the absence of global symmetries \cite{Banks:1988yz, Kallosh:1995hi, Banks:2010zn, Harlow:2018tng, Harlow:2020bee, Chen:2020ojn, Hsin:2020mfa, Yonekura:2020ino, Belin:2020jxr} and the completeness of the charge spectrum \cite{Polchinski:2003bq, Banks:2010zn, Harlow:2018tng}. Recently, the relationship between these two conditions has been clarified within the context of EFT \cite{Rudelius:2020orz, Heidenreich:2021tna}: while completeness does not follow from the absence of group-like global symmetries, it does follow from the absence of more general non-invertible global symmetries. While the context of EFT puts this result on mathematically rigorous ground, it does lose something of the spirit of the Swampland Program, and one might hope for an answer that more fully embraces the gravitational context.

\begin{figure}[h]
\centering
\includegraphics[height = 0.8 in]{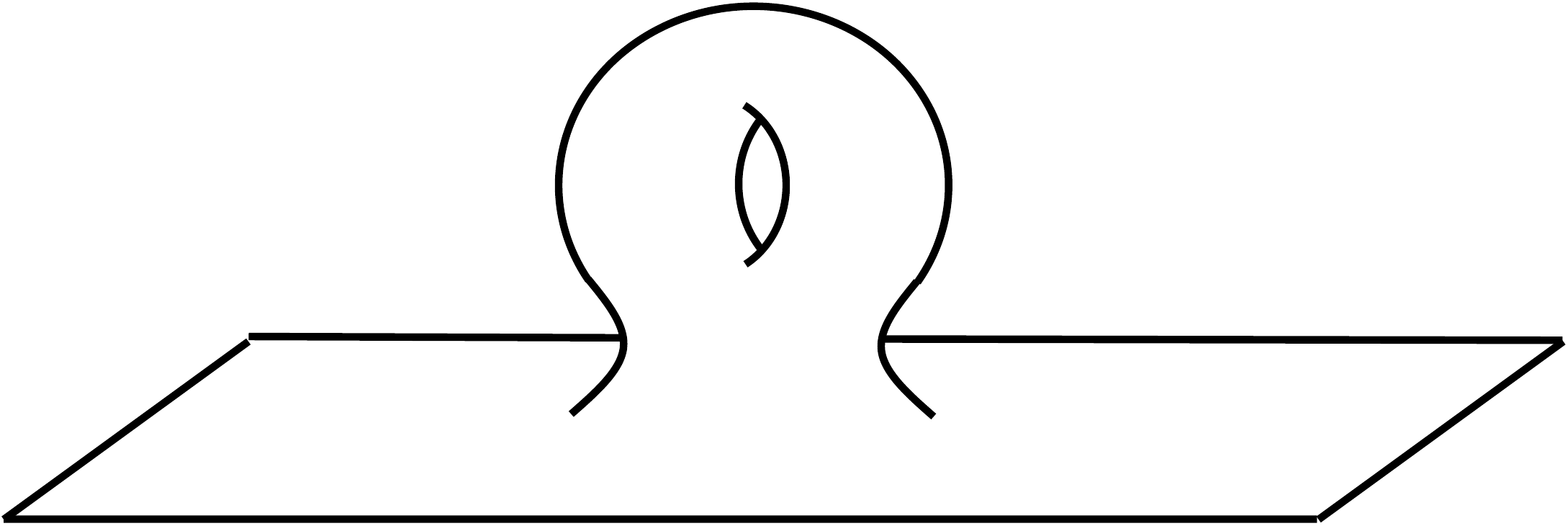}
\caption{A gravitational soliton localized in space.}
\label{fig:soliton}
\end{figure}

In this note, we extend the results of \cite{Rudelius:2020orz, Heidenreich:2021tna} by including charged gravitational solitons in the analysis. For our purposes, a \emph{gravitational soliton} is a localized configuration of a gravitational theory with nontrivial topology, as illustrated in Figure \ref{fig:soliton}, that may serve as a particle (or brane) excitation of the system \cite{Witten:1985xe, McNamara:2019rup}. As the questions we seek to address are purely kinematic, we make no attempt to find solutions to the equations of motion or stable particle states, but instead seek only finite-energy configurations. If we find a configuration carrying a certain gauge charge, then while the soliton might decay under time evolution, the end result of the decay will be a stable state of the same charge. This is enough to answer kinematic questions such as the completeness of the spectrum.

Perhaps surprisingly, we find that gravitational solitons naturally carry charges beyond those of any local excitation of the quantum fields. Correspondingly, we learn that there are symmetries of the EFT which are broken as soon as the topology of spacetime is allowed to fluctuate, even without any additional UV degrees of freedom. In fact, the effect of gravitational solitons is precisely to break the non-invertible symmetry to the maximal group-like sub-symmetry. If we further demand that this remaining group-like symmetry is broken by additional degrees of freedom, we find a complete spectrum, and thus the Completeness Hypothesis follows from the absence of group-like symmetries once gravitational solitons are taken into account.

\section{Charged Gravitational Solitons}\label{electric}

Fix a compact Lie group $G$.\footnote{The gauge group in quantum gravity is expected to be compact \cite{Polchinski:2003bq, Banks:2010zn, Harlow:2018tng}, so this is not a severe restriction.} In this section, we consider gravitational solitons in $G$ gauge theory that behave as particle excitations with electric charge. Recall that the electric charges of particles in $G$ gauge theory are described by unitary representations $\rho \in {\rm Rep}(G)$, which fuse according to the tensor product. Without gravity, the only local charged excitations in the pure gauge theory are gluons, transforming in the adjoint representation $\mathfrak{g}$. Correspondingly, the only charged local operators\footnote{By ``charged local operator," we mean a point operator attached to a Wilson line in the appropriate representation.} are those built from the field strength $F^a_{\mu \nu}$, transforming in representations built from the adjoint.

Once we include gravitational solitons, however, a new possibility arises. Consider a gravitational soliton with a nontrivial 1-cycle, as in Figure \ref{fig:soliton}, that may support a nontrivial holonomy $g \in G$ of the gauge field. The state of such a soliton may be described by a wavefunction
\beq
\psi(g) \in L^2(G),
\eeq
if we ignore other modes of the soliton. Under a gauge transformation $h$, the holonomy is conjugated, and thus the wavefunction transforms as
\beq
\psi(g) \to \psi(h g h^{-1}),
\eeq
the conjugation action of $G$ on $L^2(G)$. Thus, gravitational solitons at least provide states of every charge appearing in $L^2(G)$, and tensor products thereof.

How can we characterize this set of charges? Note that the center $Z \subset G$ is the kernel of the conjugation representation, and thus $L^2(G)$ descends to a faithful representation of the quotient group $G/Z$. By general results in the representation theory of compact groups,\footnote{See e.g. \cite[Appendix A]{Harlow:2018tng}.} we learn that every irreducible representation of $G/Z$ appears in some finite tensor product of irreducible representations appearing in $L^2(G)$. Thus, the set of charges realized by gravitational solitons includes at least every representation of $G$ in which the center acts trivially.

One might hope to produce additional charges by considering more complicated gravitational solitons, given by gauge fields on more complicated topologies. However, this is not possible, and in fact the charges realized by gravitational solitons are \emph{precisely} those representations in which the center acts trivially. In order to see this, recall that a gauge field may be reconstructed uniquely from its holonomy around all loops \cite{Barrett:1991aj, Caetano:1993zf}. Since a constant gauge transformation by $h \in Z$ acts trivially on all holonomies, it acts trivially on the gauge field, and so must act trivially in any state given by gauge fields on any topology.

Thus, we have obtained a complete characterization of the set of charges carried by gravitational solitons, as the set of representations of $G$ in which $Z$ acts trivially, or equivalently, as the set of representations of $G/Z$. In fact, this set may be characterized in another way: the center acts trivially in an irreducible representation $\rho$ if and only if the corresponding Wilson line operator $W_\rho$ has vanishing charge under the $1$-form center symmetry. Thus, gravitational solitons provide a maximal set of charged particles compatible with the preservation of the $1$-form symmetry.

Is this set of charges the same as those carried by local excitations? Put differently, can every representation $\rho$ in which the center acts trivially be found in some tensor power of the adjoint? For connected gauge groups, this is in fact the case \cite{Heidenreich:2021tna}. However, for a disconnected gauge group, this need not be the case, and the set of charges carried by solitons can be strictly larger than the set of charges built from tensor powers of the adjoint. We illustrate this possibility with the following example \cite{Heidenreich:2021tna}. 

\begin{ex}
Consider $O(2) = U(1) \rtimes \mathbb{Z}_2$ gauge theory, obtained from $U(1)$ gauge theory by gauging charge conjugation. Photons transform in the determinant representation $\bm{\mathrm{det}}$, which satisfies
\beq
\bm{\mathrm{det}} \otimes \bm{\mathrm{det}} = \bm{1},
\eeq
while the set of representations in which the center $Z = \{\pm 1\}$ acts trivially includes all representations of even charge under the identity component $U(1)$. These representations of even charge are realized by wavefunctions $\psi(g) \in L^2(O(2))$ supported on the non-identity component of $O(2)$.
\end{ex}

Now, suppose we wish to impose the swampland condition of the absence of global symmetries, by adding additional matter fields transforming in some set of representations $\{\nu_i\}$, for $1 \leq i \leq n$. In order to fully break the $1$-form center symmetry, the collection $\{\nu_i\}$ must be such that every element $h \in Z$ acts nontrivially in at least one $\nu_i$. Put differently, the direct sum
\beq
\rho = L^2(G) \oplus \nu_1 \oplus \cdots \oplus \nu_n,
\eeq
must be a faithful representation of $G$, as every element of $G \smallsetminus Z$ already acts nontrivially in $L^2(G)$. By the representation theory of compact groups, this implies that gravitational solitons and matter particles together generate a complete spectrum. Thus, the Completeness Hypothesis follows from the absence of the center symmetry in general, once the charges of gravitational solitons are included.

\section{Breaking of the Non-Invertible Symmetry and Endability}

How does this result, that completeness follows from the absence of a center symmetry, compare to the results of \cite{Rudelius:2020orz, Heidenreich:2021tna}? While we have seen that gravitational solitons preserve the center symmetry, there is an additional symmetry of the non-gravitational theory which is broken once we allow the topology of spacetime to fluctuate. This is the non-invertible 1-form symmetry, generated by topological Gukov-Witten operators. The complete breaking of this non-invertible symmetry is identified with the presence of a complete spectrum, as the non-invertible operators are able to detect the presence of charges in which the center acts trivially, yet which are absent in the pure gauge theory.

As we have seen, these are precisely the representations realized by gravitational solitons, and so we see that the non-invertible symmetry is automatically broken to the maximal group-like sub-symmetry by fluctuations of the topology. This breaking is possible because, in general, non-invertible topological operators cannot be deformed within a homology class, and need only be invariant under isotopy \cite{Chang:2018iay}, as a result of the nontrivial splitting and rejoining required to move a non-invertible operator past a handle.

As discussed in \cite{Rudelius:2020orz, Heidenreich:2021tna}, the breaking of a 1-form symmetry is correlated with the endability of line operators charged under the 1-form symmetry. In the context of $G$ gauge theory, this is the endability of Wilson line operators $W_\rho$ on charged local operators $\mathcal{O}_\rho$. As we have seen, the only local operators in $G$ gauge theory are built from the adjoint, and so there are no local operators on which a Wilson line $W_\rho$ can end if $\rho$ does not appear in any tensor power of the adjoint. This is consistent with the preservation of the non-invertible symmetry in the non-gravitational theory.

At first glance, it might appear that nothing changes once gravity is made dynamical, as there do not appear to be any additional charged local operators. This presents a puzzle: our analysis of charged gravitational solitons suggests the non-invertible symmetry is broken, and yet there are no local operators on which the corresponding Wilson lines may end.

\begin{figure}[h]
\centering
\includegraphics[height = 1 in]{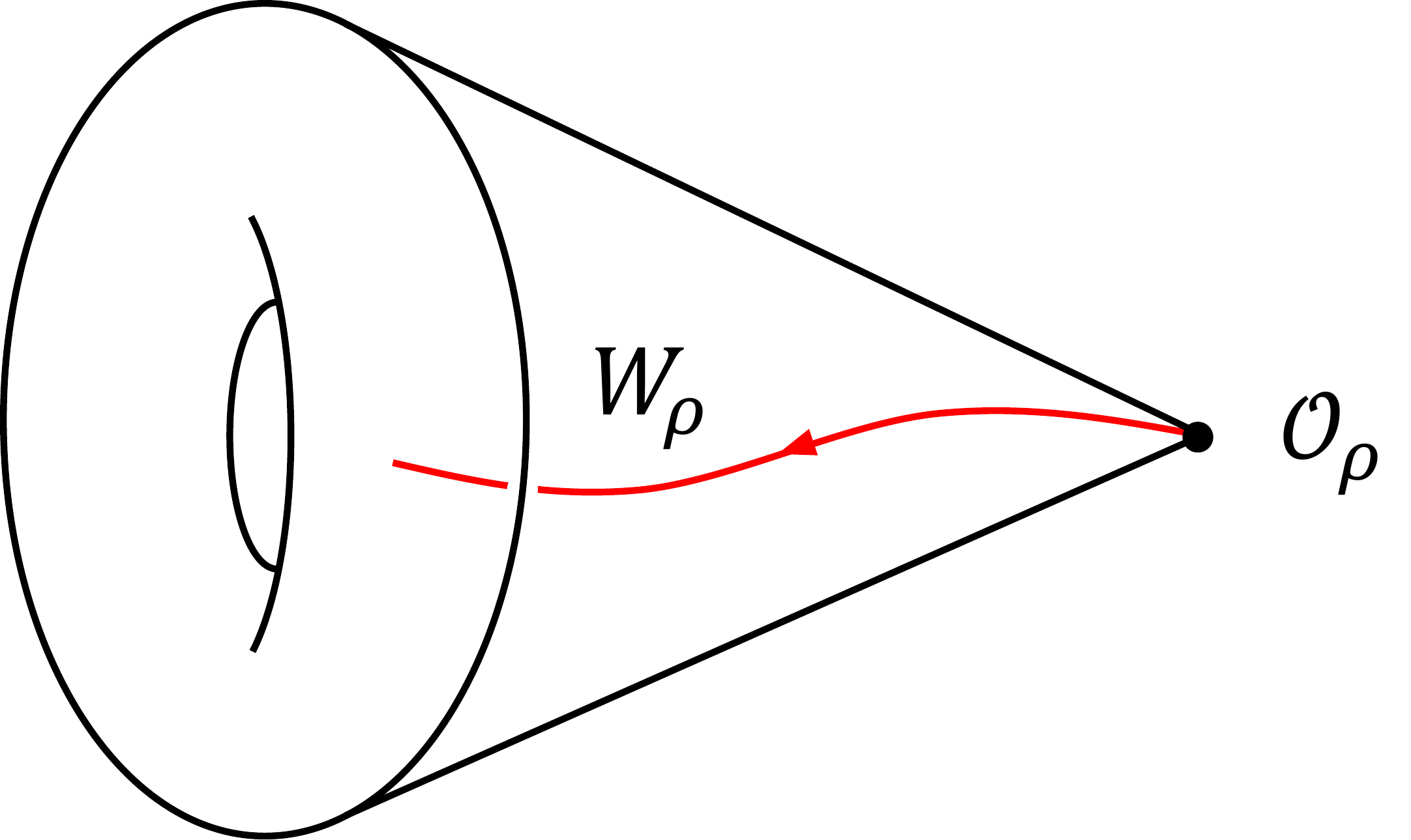}
\caption{Charged gravitational solitons are created by charged operators at cone points.}
\label{fig:cone}
\end{figure}

The resolution is that, as the new charged states are gravitational solitons with nontrivial topology, the operators that create such states from the vacuum must be inserted at conical singularities, as illustrated in Figure \ref{fig:cone}, rather than at smooth points in spacetime. More precisely, given a gravitational soliton with some spatial topology that is asymptotic to $\mathbb{R}^{d - 1}$, let $\mathcal{M}$ be the 1-point compactification. The operator that creates the soliton must sit at the cone point of
\beq
{\rm C} \mathcal{M} = \frac{\mathcal{M} \times \mathbb{R}_+}{(m, 0) \sim (m', 0)},
\eeq
so that its insertion induces a change of the topology. Such operators are not valid local operators in quantum field theory, but must be considered on the same footing in quantum gravity, where the topology is dynamical.

\vspace{6pt}

For an alternative perspective, note that such an operator inserted at the cone point may be thought of as creating a state on the closed manifold $\mathcal{M}$. If the operator is attached to a Wilson line $W_\rho$, this Wilson line will extend out to puncture $\mathcal{M}$ at a point, and so this state lives in a defect Hilbert space $\mathcal{H}_\rho(\mathcal{M})$ of states of charge $\rho$ on $\mathcal{M}$, whereas in quantum field theory we were restricted to the case $\mathcal{M} = S^{d-1}$, corresponding to local operators at a smooth point.

\section{Wormholes and the Adjoint Subcategory}

Having analyzed the case of gravitational solitons with electric charge under $G$ gauge theory, let us now turn to more general notions of charged particle, such as dyonic charges in four dimensions or anyonic charges in three dimensions. Most generally, we consider particle charges described by a semisimple tensor category $\mathcal{C}$ of line operators.\footnote{See \cite{Etingof} for a textbook account of tensor categories.} We work in at least three spacetime dimensions, so that $\mathcal{C}$ is braided.

To study these cases, let us seek a more invariant description of the set of charges realized by gravitational solitons in $G$ gauge theory. As we saw, the charges of gravitational solitons are generated by the irreducible representations in $L^2(G)$, acted on by conjugation. The structure of this representation is well-understood, and is described by the Peter-Weyl theorem as
\beq
L^2(G) = \bigoplus_\rho \rho \otimes \overline{\rho},
\eeq
where $\rho$ runs over a complete set of irreducible representations.

\begin{figure}[h]
\centering
\includegraphics[height = 1 in]{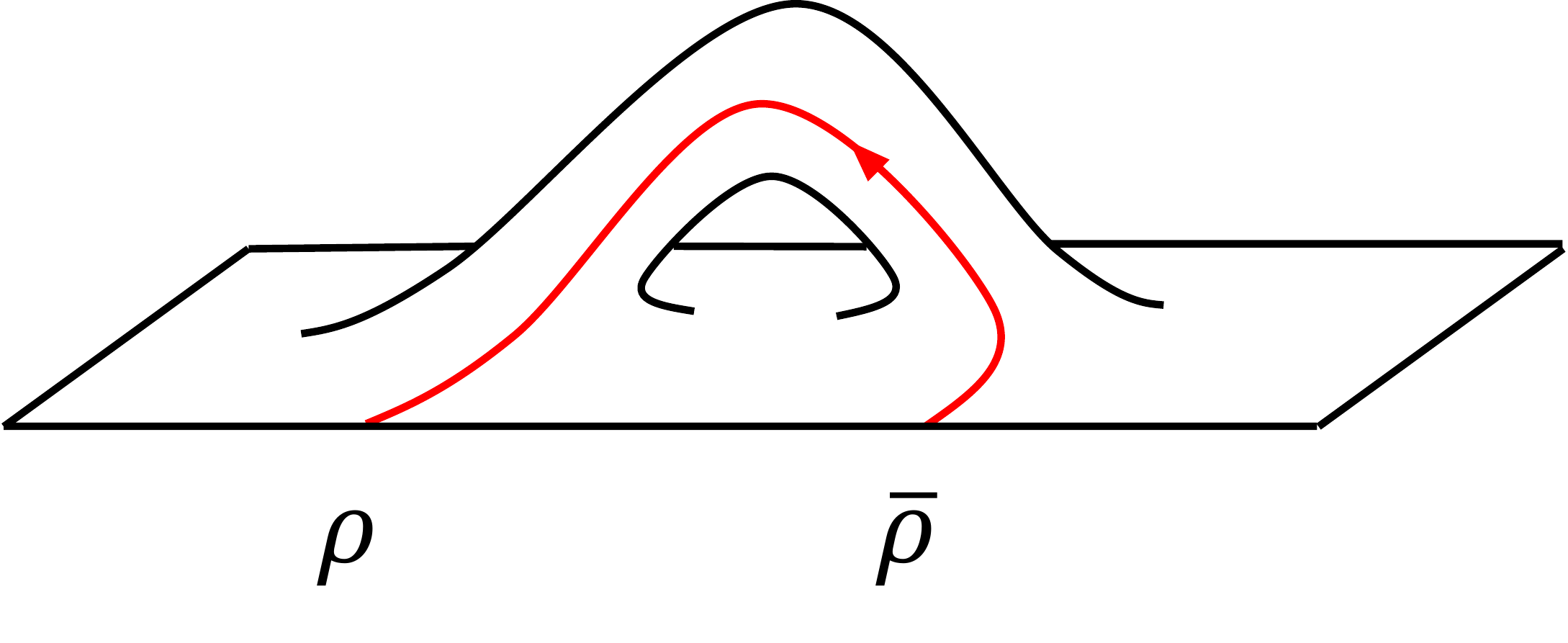}
\vspace{-6pt}
\caption{A wormhole threaded with electric flux.}
\label{fig:wormhole}
\end{figure}

This description is quite evocative, and suggests the following interpretation. One way to produce a gravitational soliton is to consider a wormhole connecting two nearby points in space, through which some electric gauge flux is flowing. If the flux is described by a representation $\rho$, then one end of the wormhole will behave as a particle of charge $\rho$, while the other behaves as a particle of charge $\overline{\rho}$, as illustrated in Figure \ref{fig:wormhole}.

\vspace{12pt}

If each end of the wormhole were free to move about as an independent particle, we would have a complete spectrum, as would be the case if the Cobordism Conjecture \cite{McNamara:2019rup} were satisfied and the wormhole could pinch off into two independent ends. Without imposing the Cobordism Conjecture, the two ends are not independent particles, but we may bring the ends together to form a gravitational soliton of charge $\rho \otimes \overline{\rho}$, in line with our previous analysis.

This picture, in which we construct solitons by bringing the ends of a wormhole together, applies in much more generality than electric charges under a gauge group. If our particle charges are described by a braided semisimple tensor category $\mathcal{C}$, the above argument tells us that the charges realized by gravitational solitons include at least the \emph{adjoint subcategory} $\mathcal{C}_{\rm ad}$, defined to be the smallest tensor Serre subcategory of $\mathcal{C}$ containing all objects $\rho \otimes \overline{\rho}$, as $\rho$ ranges over the simple objects of $\mathcal{C}$.

In fact, the general theory of tensor categories implies that the charges of gravitational solitons are \emph{exactly} given by the adjoint subcategory $\mathcal{C}_{\rm ad}$, as it is the maximal subcategory compatible with the preservation of a group-like 1-form symmetry. By \cite[Definition 4.14.2]{Etingof}, the braided tensor category $\mathcal{C}$ has a universal grading,
\beq
\mathcal{C} = \bigoplus_{u \in U_\mathcal{C}} \mathcal{C}_u,
\eeq
where $U_\mathcal{C}$ is an abelian group, the \emph{universal grading group}, such that the trivially graded component is the adjoint subcategory $\mathcal{C}_1 = \mathcal{C}_{\rm ad}$, and such that $\mathcal{C}_u$ is irreducible as a module over $\mathcal{C}_{\rm ad}$. To connect to our previous discussion, let us consider the following example.

\begin{ex}
Let $\mathcal{C} = {\rm Rep}(G)$, the tensor category of unitary representations of a compact Lie group $G$. Let $Z$ be the center of $G$. Then the adjoint subcategory is
\beq
\mathcal{C}_{\rm ad} = {\rm Rep}(G/Z),
\eeq
the category of representations of $G$ in which $Z$ acts trivially. The universal grading group is
\beq
U_\mathcal{C} = Z^\vee
\eeq
the Pontryagin dual of $Z$, and the $u$-graded subcategories, for $u : Z \to U(1)$, are
\beq
\mathcal{C}_u = \left\{ \rho \in {\rm Rep}(G)\ |\ \text{$Z$ acts in $\rho$ by $u$} \right\}.
\eeq
\end{ex}

Motivated by this example, we might expect that a theory whose charges are described by $\mathcal{C}$ might have a $U_\mathcal{C}^\vee$ 1-form symmetry, by analogy to the 1-form center symmetry of $G$-gauge theory. This is in fact the case: by \cite[Proposition 4.14.3]{Etingof}, the delooping $B U_\mathcal{C}^\vee$ acts as a symmetry of the category $\mathcal{C}$ of particle charges, which is equivalent to saying that $U_\mathcal{C}^\vee$ is a 1-form symmetry. More concretely, we may define codimension-2 symmetry operators for each element of $U_\mathcal{C}^\vee$ by their linking with the line operators representing probe particles.

Thus, we see that the results of our previous analysis hold in far more generality. Gravitational solitons provide particles with all charges not protected by a group-like 1-form symmetry. If we impose the absence of global symmetries by adding additional particles to fully break the symmetry, the irreducibility of $\mathcal{C}_u$ as a module over $\mathcal{C}_{\rm ad}$ implies that gravitational solitons and the additional particles generate a complete spectrum. Thus, completeness of the particle spectrum follows in general from the absence of global symmetries and the presence of gravitational solitons.

\section{Higher Brane Charges}

So far, we have only considered gravitational solitons that are localized at a point in space, which serve as particle excitations of the gravitational theory. However, quantum theories of gravity contain not only particles, but also extended objects of various dimensions, such as strings and branes, and we would like to describe the higher brane charges carried by gravitational solitons as well.

In principle, one should consider the entire spectrum of brane charges of different dimension at once, to capture effects such as higher-group symmetry \cite{Cordova:2018cvg} or the solubility of different branes in each other \cite{Heidenreich:2020pkc}. In $d$ dimensions, these charges are described by a semisimple $d$-category $\mathcal{C}$,\footnote{See \cite{douglas2018fusion, JohnsonFreydTalk} for basic definitions and constructions in such categories.} whose $k$-morphisms describe charges for objects of codimension $k$ in spacetime. Thus, the objects of $\mathcal{C}$ are vacua of the theory, the 1-morphisms are domain walls, and so on. If we only care about charges of objects of codimension $k \geq k_0$ in a certain vacuum $V$, we may restrict attention to $\Omega_V^{k_0} \mathcal{C}$, the $k_0$-fold looping of $\mathcal{C}$ at $V$, a rigid $k_0$-tuply monoidal semisimple $(d - k_0)$-category.

As we saw in the previous section, the particle charges of gravitational solitons are described by the adjoint subcategory, generated by fusions $\rho \otimes \overline{\rho}$ of simple objects with their duals. As far as we are aware, the theory of the adjoint subcategory and universal grading has not been extended to the $d$-categorical case, but we might hope for a similar picture to hold.

To make this more precise, for $k > 0$, consider the set $\pi_k \mathcal{C}$ of Schur equivalence classes of simple $k$-morphisms, i.e., codimension-$k$ brane charges. We define the \emph{adjoint subset} $( \pi_k \mathcal{C})_{\rm ad}$ to be the set of all simple $k$-morphisms appearing in fusions $\rho \otimes \overline{\rho}$ of simple $k$-morphisms with their duals. Further, we define $U_{\pi_k \mathcal{C}}$ to be the universal grading on $\pi_k \mathcal{C}$ compatible with the fusion. We expect to find gravitational solitons of codimension $k$ with charges in $(\pi_k \mathcal{C})_{\rm ad}$, as well as a $U_{\pi_k \mathcal{C}}^\vee$ $(d - k)$-form symmetry, whose breaking implies completeness.

\begin{figure}[h]
\centering
\includegraphics[height = 0.8 in]{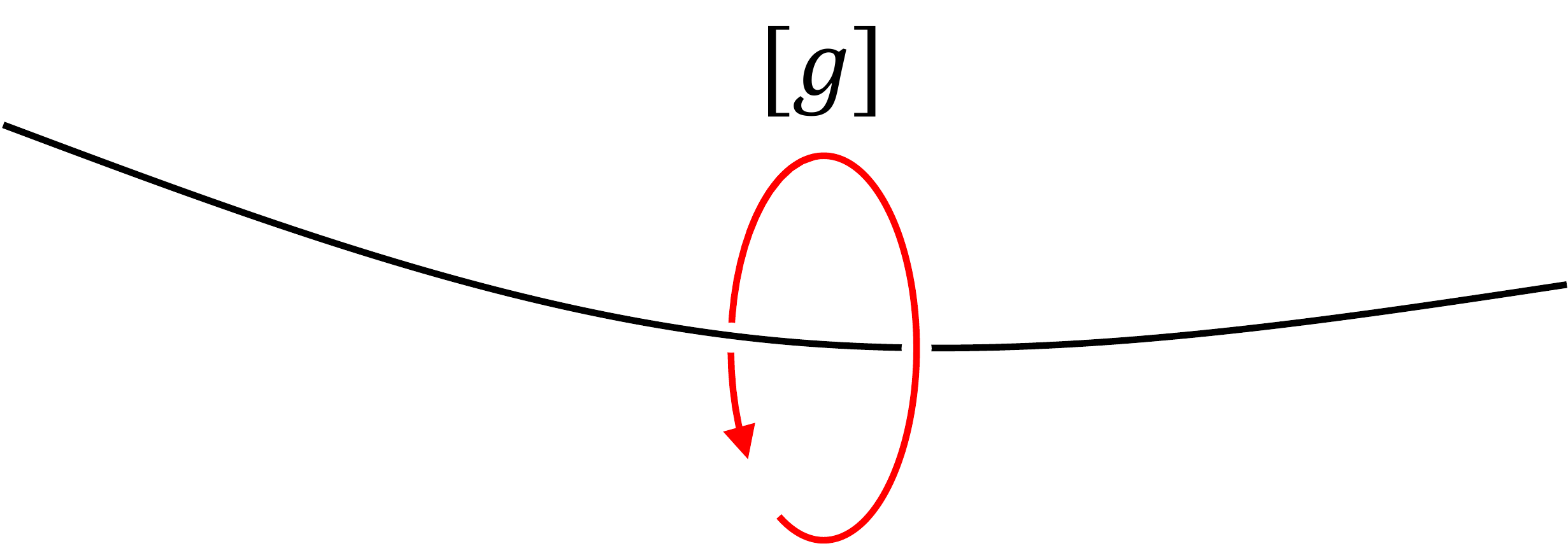}
\caption{A codimension-$2$ vortex with holonomy $[g]$.}
\label{fig:vortex}
\end{figure}

To illustrate this, we consider an example which has been the subject of recent discussion in the literature \cite{Rudelius:2020orz, Heidenreich:2021tna, Dierigl:2020lai}: the charges of codimension-2 vortices defined by the holonomy of a discrete gauge group $\Gamma$, as illustrated in Figure \ref{fig:vortex}.

\vspace{6pt}

\begin{ex} Let $\Gamma$ be a discrete group. Consider the simply-connected semisimple $d$-category $\mathcal{C}$ defined by
\beq
\Omega^2 \mathcal{C} = Z\left({\rm Vect}_{d - 2}[\Gamma]\right),
\eeq
where ${\rm Vect}_{d-2}[\Gamma]$ is the monoidal $(d - 2)$-category of $\Gamma$-graded $(d-2)$-vector spaces, and $Z(-)$ denotes the Drinfeld center. For $d \geq 4$,\footnote{In $d = 3$, vortices are particles, and so we must consider fusions of vortices with electric particles as well, labeled by pairs $([g], \rho)$ of a conjugacy class $[g]$ and a representation $\rho$ of the stabilizer of $g$ (see e.g. \cite{Dijkgraaf:1989pz, Kitaev:1997wr}).} the set of Schur equivalence classes of simple $2$-morphisms in $\mathcal{C}$ is
\beq
\pi_2 \mathcal{C} = {\rm conj}(\Gamma),
\eeq
the set of conjugacy classes in $\Gamma$, with fusion given by the fusion of conjugacy classes \cite{Dijkgraaf:1989hb}. Concretely, a conjugacy class $[g]$ corresponds to a vortex defined by a discrete holonomy $g \in \Gamma$, defined only up to gauge transformation, i.e., conjugation.

We compute
\beq
[g] \otimes \overline{[g]} = [g] \otimes [g^{-1}] \supset [g h g^{-1} h^{-1}],
\eeq
for any $h \in \Gamma$, and thus we have that
\beq
(\pi_2 \mathcal{C})_{\rm ad} = {\rm conj}_\Gamma[\Gamma, \Gamma],
\eeq
the set of $\Gamma$-conjugacy classes in the commutator subgroup $[\Gamma, \Gamma]$. The universal grading group is
\beq
U_{\pi_2 \mathcal{C}} = \Gamma / [\Gamma, \Gamma] = \Gamma_{\rm ab},
\eeq
the abelianization of $\Gamma$, where the $a$-graded subset for $a \in \Gamma_{\rm ab}$ is given by the conjugacy classes that project to $a$ in the abelianization.

\end{ex}

\begin{figure}[h]
\centering
\includegraphics[height = 1 in]{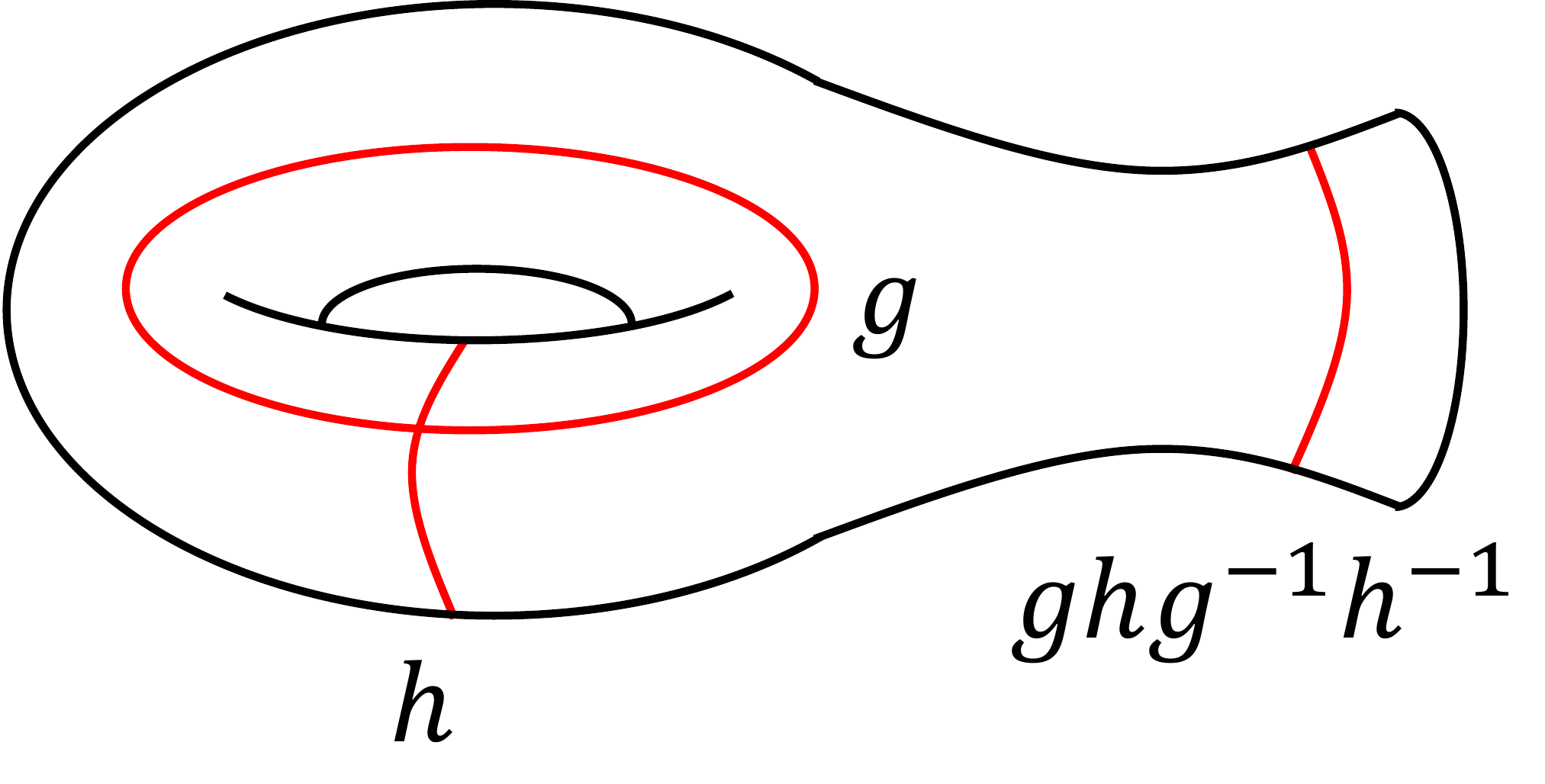}
\caption{A vortex in the commutator subgroup.}
\label{fig:commutator}
\end{figure}

The upshot of this example is that gravitational solitons are expected to produce vortices for any $\Gamma$-conjugacy class in the commutator subgroup $[\Gamma, \Gamma]$, and we expect a $\Gamma_{\rm ab}^\vee$ $(d - 2)$-form symmetry. Indeed, both are the case! Note that
\beq
\Gamma_{\rm ab}^\vee = {\rm Hom}(\Gamma, U(1)),
\eeq
and so we may identify the $\Gamma_{\rm ab}^\vee$ symmetry as the symmetry generated by invertible, topological Wilson lines $W_\rho$, labeled by 1-dimensional unitary representations $\rho$ of $\Gamma$ \cite{Rudelius:2020orz, Heidenreich:2021tna}. A gravitational soliton realizing a conjugacy class $[g h g^{-1} h^{-1}]$ is given by a punctured torus with discrete holonomies $g$ and $h$ around the two 1-cycles, as illustrated in Figure \ref{fig:commutator}.

\vspace{12pt}
\vspace{-6pt}

While gravitational solitons produced by codimension-$k$ wormholes carry charges in $(\pi_k \mathcal{C})_{\rm ad}$, this is not the most general type of gravitational soliton we can construct, and a full analysis would involve not just wormholes, but also solitons constructed by surgeries along higher dimensional manifolds. We illustrate this possibility with the following example.

\begin{ex}

Consider axion electrodynamics in four dimensions,\footnote{The higher-group symmetries of axion electrodynamics have been a subject of recent study \cite{Heidenreich:2021tna, Heidenreich:2020pkc, Hidaka:2020iaz, Hidaka:2020izy, Brennan:2020ehu, Hidaka:2021mml}.} defined by a $U(1)$ gauge field $A$ and a compact scalar $\phi$, together with an axion coupling
\beq
\mathcal{L} \supset \phi\ F \wedge F.
\eeq
While the $U(1)$ shift symmetry of $\phi$ is broken by the axion coupling, there are no instantons coupled to $\phi$ in the EFT, as there are no $U(1)$ gauge instantons localized in $\mathbb{R}^4$ in continuum quantum field theory.\footnote{See \cite[Section 8.2]{Heidenreich:2021tna} for similar discussion. Of course, point-size abelian instantons are familiar in string theory and field theories on non-commutative spacetimes. In fact, these instantons are likely equivalent to gravitational solitons \cite{Braden:1999zp}.}

However, we may construct a gravitational soliton with nonzero instanton number as follows. Excise from $\mathbb{R}^4$ an embedded copy of $S^1 \times D^3$, and glue in a copy of $D^2 \times S^2$, to obtain a manifold diffeomorphic to $\mathbb{R}^4 \# (S^2 \times S^2)$. To produce a configuration with nonzero instanton number, we simply place magnetic flux of $F$ on each $S^2$. The resulting configuration is a solitonic version of the monopole loops studied in \cite{Fan:2021ntg}, just as the wormholes considered previously are solitonic versions of particle-antiparticle pairs.

\end{ex}

It would be quite interesting to give a complete description of the charges realized by gravitational solitons of every dimension. Such a description might help illuminate the proper generalization of the adjoint subcategory to the case of a general semisimple $d$-category.

\section{Discussion}

We have seen that the consideration of gravitational solitons leads to a tremendous simplification of the question of completeness of the spectrum of gauge charges. While the question of completeness in EFT is quite complicated, involving non-invertible symmetries \cite{Rudelius:2020orz, Heidenreich:2021tna} and non-abelian charge structures \cite{Dierigl:2020lai}, the question of completeness in quantum gravity reduces to the breaking of an abelian, group-like symmetry.

In particular, this observation resolves a reasonable objection to the claim that completeness follows from the Cobordism Conjecture \cite{McNamara:2019rup}. Namely, the symmetries arising from the cobordism groups of quantum gravity are always abelian and group-like, and so it may have been difficult to imagine that their absence could imply completeness in general. The results of this note show that such a simplification is natural, and indeed the cobordism groups of quantum gravity measure the symmetries left unbroken even after the topology is allowed to fluctuate.\footnote{Note the tension between this result and the claim of \cite{Yonekura:2020ino} that \emph{all} higher-form symmetries are automatically broken by fluctuations of the topology.}

Let us emphasize that the effect described in this note is a direct result of the fact that in quantum gravity, we cannot assume that spacetime is topologically trivial at short distance scales, so that nontrivial topologies supporting holonomies and fluxes can provide particle and brane excitations beyond those of any local quantum field. Interestingly, the same is true of lattice systems, where the nontrivial topology at the lattice scale allows for degrees of freedom with exactly the same charges \cite[Appendix B]{Casini:2020rgj} as we found for gravitational solitons. This hints at a deep connection between quantum gravity and lattice systems \cite{WenStringsTalk}, and could perhaps shed light on the appearance of cobordism in both \cite{McNamara:2019rup, Freed:2016rqq}.

Finally, a shortcoming of the perspective advocated in this note is that, generically, we would expect the masses of gravitational solitons to be near the Planck scale, and so their relevance for low-energy physics may be somewhat suspect. While this expectation is valid semiclassically, it is conceivable that strong quantum corrections could bring their masses down. In fact, once quantum gravity becomes strongly-interacting, the distinction between gravitational soliton and localized particle becomes blurred, and what appears to be a gravitational soliton in one duality frame might be a local particle in another. Understanding this possibility is necessary if we want to understand the full implications of gravitational solitons for the Swampland Program, and in particular whether they can provide the light states demanded by the Weak Gravity Conjecture \cite{Arkani-Hamed:2006emk} and the Swampland Distance Conjecture \cite{Ooguri:2006in}. 

\section*{Acknowledgments}

We thank Sergio Cecotti, Ben Heidenreich, Miguel Montero, Matthew Reece, Tom Rudelius, Cumrun Vafa, and Irene Valenzuela for many discussions and closely related collaborations. We thank Theo Johnson-Freyd, Zohar Komargodski, and Sahand Seifnashri for discussions on semisimple higher categories. We thank Matthew Reece for comments on the draft. We thank the
Simons Center for Geometry and Physics, where part of this work was completed during the 2019 and 2021 summer workshops.

The research of JM is supported by the National Science Foundation Graduate Research Fellowship Program under Grant No.~DGE1745303. Any opinions, findings, and conclusions or recommendations expressed in this
material are those of the author and do not necessarily reflect the views of the National
Science Foundation.


\begin{thebibliography}{14}

\bibitem{Hawking:1978pog}
S.~W.~Hawking,
``Space-Time Foam,''
Nucl. Phys. B \textbf{144}, 349-362 (1978).

\bibitem{Brennan:2017rbf}
T.~D.~Brennan, F.~Carta and C.~Vafa,
``The String Landscape, the Swampland, and the Missing Corner,''
\emph{PoS} TASI2017, \textbf{015} (2017),
arXiv:1711.00864 [hep-th].

\bibitem{Palti:2019pca}
E.~Palti,
``The Swampland: Introduction and Review,''
Fortsch. Phys. \textbf{67}, no.6, 1900037 (2019),
arXiv:1903.06239 [hep-th].

\bibitem{vanBeest:2021lhn}
M.~van Beest, J.~Calder\'on-Infante, D.~Mirfendereski and I.~Valenzuela,
``Lectures on the Swampland Program in String Compactifications,''
arXiv:2102.01111 [hep-th].

\bibitem{Grana:2021zvf}
M.~Gra\~na and A.~Herr\'aez,
``The Swampland Conjectures: A bridge from Quantum Gravity to Particle Physics,''
arXiv:2107.00087 [hep-th].

\bibitem{Banks:1988yz}
T.~Banks and L.~J.~Dixon,
``Constraints on String Vacua with Space-Time Supersymmetry,''
Nucl. Phys. B \textbf{307}, 93-108 (1988).

\bibitem{Kallosh:1995hi}
R.~Kallosh, A.~D.~Linde, D.~A.~Linde and L.~Susskind,
``Gravity and global symmetries,''
Phys. Rev. D \textbf{52}, 912-935 (1995),
arXiv:hep-th/9502069 [hep-th].

\bibitem{Banks:2010zn}
T.~Banks and N.~Seiberg,
``Symmetries and Strings in Field Theory and Gravity,''
Phys. Rev. D \textbf{83}, 084019 (2011),
arXiv:1011.5120 [hep-th].

\bibitem{Harlow:2018tng}
D.~Harlow and H.~Ooguri,
Commun. Math. Phys. \textbf{383}, no.3, 1669-1804 (2021),
arXiv:1810.05338 [hep-th].

\bibitem{Harlow:2020bee}
D.~Harlow and E.~Shaghoulian,
``Global symmetry, Euclidean gravity, and the black hole information problem,''
JHEP \textbf{04}, 175 (2021),
arXiv:2010.10539 [hep-th].

\bibitem{Chen:2020ojn}
Y.~Chen and H.~W.~Lin,
``Signatures of global symmetry violation in relative entropies and replica wormholes,''
JHEP \textbf{03}, 040 (2021),
arXiv:2011.06005 [hep-th].

\bibitem{Hsin:2020mfa}
P.~S.~Hsin, L.~V.~Iliesiu and Z.~Yang,
``A violation of global symmetries from replica wormholes and the fate of black hole remnants,''
arXiv:2011.09444 [hep-th].

\bibitem{Yonekura:2020ino}
K.~Yonekura,
``Topological violation of global symmetries in quantum gravity,''
arXiv:2011.11868 [hep-th].

\bibitem{Belin:2020jxr}
A.~Belin, J.~De Boer, P.~Nayak and J.~Sonner,
``Charged Eigenstate Thermalization, Euclidean Wormholes and Global Symmetries in Quantum Gravity,''
arXiv:2012.07875 [hep-th].

\bibitem{Polchinski:2003bq}
J.~Polchinski,
Int. J. Mod. Phys. A \textbf{19S1}, 145-156 (2004),
arXiv:hep-th/0304042 [hep-th].

\bibitem{Rudelius:2020orz}
T.~Rudelius and S.~H.~Shao,
``Topological Operators and Completeness of Spectrum in Discrete Gauge Theories,''
JHEP \textbf{12}, 172 (2020),
arXiv:2006.10052 [hep-th].

\bibitem{Heidenreich:2021tna}
B.~Heidenreich, J.~Mcnamara, M.~Montero, M.~Reece, T.~Rudelius and I.~Valenzuela,
``Non-Invertible Global Symmetries and Completeness of the Spectrum,''
arXiv:2104.07036 [hep-th].

\bibitem{Witten:1985xe}
E.~Witten,
``Global Gravitational Anomalies,''
Commun. Math. Phys. \textbf{100}, 197 (1985).

\bibitem{McNamara:2019rup}
J.~McNamara and C.~Vafa,
``Cobordism Classes and the Swampland,''
arXiv:1909.10355 [hep-th].

\bibitem{Barrett:1991aj}
J.~W.~Barrett,
``Holonomy and path structures in general relativity and Yang-Mills theory,''
Int. J. Theor. Phys. \textbf{30}, 1171-1215 (1991).

\bibitem{Caetano:1993zf}
A.~Caetano and R.~F.~Picken,
``An Axiomatic definition of holonomy,''
Int. J. Math. \textbf{5} 835 (1994).

\bibitem{Chang:2018iay}
C.~M.~Chang, Y.~H.~Lin, S.~H.~Shao, Y.~Wang and X.~Yin,
``Topological Defect Lines and Renormalization Group Flows in Two Dimensions,''
JHEP \textbf{01}, 026 (2019),
arXiv:1802.04445 [hep-th].

\bibitem{Etingof}
P.~Etingof, S.~Gelaki, D.~Nikshych and V.~Ostrik,
``Tensor Categories,"
Mathematical Surveys and Monographs \textbf{205}, American Mathematical Society, U.S.A. (2016).

\bibitem{Cordova:2018cvg}
C.~C\'ordova, T.~T.~Dumitrescu and K.~Intriligator,
``Exploring 2-Group Global Symmetries,''
JHEP \textbf{02}, 184 (2019),
arXiv:1802.04790 [hep-th].

\bibitem{Heidenreich:2020pkc}
B.~Heidenreich, J.~McNamara, M.~Montero, M.~Reece, T.~Rudelius and I.~Valenzuela,
``Chern-Weil Global Symmetries and How Quantum Gravity Avoids Them,''
arXiv:2012.00009 [hep-th].

\bibitem{douglas2018fusion}
C.~Douglas and D.~Reutter,
``Fusion 2-categories and a state-sum invariant for 4-manifolds,"
arXiv:1812.11933 [math.QA]

\bibitem{JohnsonFreydTalk}
T.~Johnson-Freyd,
``Semisimple Higher Categories,"
Western Hemisphere Colloquium on Geometry and Physics (Jul. 26, 2021),
available at \url{https://youtu.be/-20-lxcCOiQ}

\bibitem{Dierigl:2020lai}
M.~Dierigl and J.~J.~Heckman,
``Swampland cobordism conjecture and non-Abelian duality groups,''
Phys. Rev. D \textbf{103}, no.6, 066006 (2021),
arXiv:2012.00013 [hep-th].

\bibitem{Dijkgraaf:1989pz}
R.~Dijkgraaf and E.~Witten,
``Topological Gauge Theories and Group Cohomology,''
Commun. Math. Phys. \textbf{129}, 393 (1990).

\bibitem{Kitaev:1997wr}
A.~Y.~Kitaev,
``Fault tolerant quantum computation by anyons,''
Annals Phys. \textbf{303}, 2-30 (2003),
arXiv:quant-ph/9707021 [quant-ph].

\bibitem{Dijkgraaf:1989hb}
R.~Dijkgraaf, C.~Vafa, E.~P.~Verlinde and H.~L.~Verlinde,
``The Operator Algebra of Orbifold Models,''
Commun. Math. Phys. \textbf{123}, 485 (1989).


\bibitem{Hidaka:2020iaz}
Y.~Hidaka, M.~Nitta and R.~Yokokura,
``Higher-form symmetries and 3-group in axion electrodynamics,''
Phys. Lett. B \textbf{808}, 135672 (2020),
arXiv:2006.12532 [hep-th].

\bibitem{Hidaka:2020izy}
Y.~Hidaka, M.~Nitta and R.~Yokokura,
``Global 3-group symmetry and 't Hooft anomalies in axion electrodynamics,''
JHEP \textbf{01}, 173 (2021),
arXiv:2009.14368 [hep-th].

\bibitem{Brennan:2020ehu}
T.~D.~Brennan and C.~Cordova,
``Axions, Higher-Groups, and Emergent Symmetry,''
arXiv:2011.09600 [hep-th].

\bibitem{Hidaka:2021mml}
Y.~Hidaka, M.~Nitta and R.~Yokokura,
``Topological axion electrodynamics and 4-group symmetry,''
arXiv:2107.08753 [hep-th].

\bibitem{Braden:1999zp}
H.~W.~Braden and N.~A.~Nekrasov,
``Space-time foam from noncommutative instantons,''
Commun. Math. Phys. \textbf{249}, 431-448 (2004),
arXiv:hep-th/9912019 [hep-th].

\bibitem{Fan:2021ntg}
J.~Fan, K.~Fraser, M.~Reece and J.~Stout,
``Axion Mass from Magnetic Monopole Loops,''
arXiv:2105.09950 [hep-ph].

\bibitem{Casini:2020rgj}
H.~Casini, M.~Huerta, J.~M.~Magan and D.~Pontello,
``Entropic order parameters for the phases of QFT,''
JHEP \textbf{04}, 277 (2021),
arXiv:2008.11748 [hep-th].

\bibitem{WenStringsTalk}
X.-G.~Wen,
``Review on Swampland in Condensed Matter Physics,"
Strings 2021, ICTP-SAIFR (Jul. 1, 2021),
available at \url{https://youtu.be/Jf6XrzeIVzM}

\bibitem{Freed:2016rqq}
D.~S.~Freed and M.~J.~Hopkins,
``Reflection positivity and invertible topological phases,''
Geom. Topol. \textbf{25}, 1165-1330 (2021),
arXiv:1604.06527 [hep-th].

\bibitem{Arkani-Hamed:2006emk}
N.~Arkani-Hamed, L.~Motl, A.~Nicolis and C.~Vafa,
``The String landscape, black holes and gravity as the weakest force,''
JHEP \textbf{06}, 060 (2007),
arXiv:hep-th/0601001 [hep-th].

\bibitem{Ooguri:2006in}
H.~Ooguri and C.~Vafa,
``On the Geometry of the String Landscape and the Swampland,''
Nucl. Phys. B \textbf{766}, 21-33 (2007),
arXiv:hep-th/0605264 [hep-th].

\end{thebibliography}
\end{document}